\documentclass{iau} 
\usepackage{graphicx}
\usepackage{natbib}
\usepackage{hyperref}

\newcommand{\msun}{\mbox{M$_{\odot}$}}

\newcommand{\lsun}{\mbox{L$_{\odot}$}}
\newcommand{\zsun}{\mbox{Z$_{\odot}$}}

\providecommand{\sun}{_\text{\sun}}

	\title[Introducing the ASSESS project] 
	{Introducing the ASSESS project: Episodic Mass Loss in Evolved Massive Stars - Key to
Understanding the Explosive Early Universe}
	
	\author[Bonanos \textit{et al.}]   
	{	A.Z. Bonanos$^{1}$,
	G. Maravelias$^{1,3}$,
		M. Yang$^{1,5}$,
		F. Tramper$^{4}$,
		S. de Wit$^{1,2}$,
		E. Zapartas$^{1}$,
		K. Antoniadis$^{1,2}$,
		E. Christodoulou$^{1,2}$,
		G. Munoz-Sanchez$^{1,2}$
		}
	
	\affiliation{$^1$IASSARS, National Observatory of Athens, Athens, Greece
		\\[\affilskip]
		$^{2}$National and Kapodistrian University of Athens, Athens, Greece
		\\[\affilskip]
		$^3$Institute of Astrophysics, FORTH, Heraklion, Greece
		\\[\affilskip]
		$^4$Institute of Astronomy, KU Leuven, Belgium
		\\[\affilskip]
		$^5$Key Laboratory of Space Astronomy and Technology, National Astronomical Observatories, Chinese Academy of Sciences, Beijing, People’s Republic of China
		\\[\affilskip]
	}
	
	\pubyear{2022}
	\volume{361}  
	\jname{Massive Stars Near and Far}
	\editors{N.~St-Louis, J.~S.~Vink \& J.~Mackey, eds.}
\begin{document}
	
	\maketitle
	
	\begin{abstract}
	
Episodic mass loss is not understood theoretically, neither accounted for in state-of-the-art models of stellar evolution, which has far-reaching consequences for many areas of astronomy. We introduce the ERC-funded ASSESS project (2018-2024), which aims to determine whether episodic mass loss is a dominant process in the evolution of the most massive stars, by conducting the first extensive, multi-wavelength survey of evolved massive stars in the nearby Universe. It hinges on the fact that mass-losing stars form dust and are bright in the mid-infrared. We aim to derive physical parameters of $\sim$1000 dusty, evolved massive stars in $\sim$25 nearby galaxies and estimate the amount of ejected mass, which will constrain evolutionary models, and quantify the duration and frequency of episodic mass loss as a function of metallicity. The approach involves applying machine-learning algorithms to select dusty, luminous targets from existing multi-band photometry of nearby galaxies. We present the first results of the project, including the machine-learning methodology for target selection and results from our spectroscopic observations so far. The emerging trend for the ubiquity of episodic mass loss, if confirmed, will be key to understanding the explosive early Universe and will have profound consequences for low-metallicity stars, reionization, and the chemical evolution of galaxies.
	\end{abstract}

\keywords{
Stars: massive -- Stars: mass loss --  Stars: evolution  -- Stars: fundamental parameters -- Supergiants -- Surveys}

	\firstsection 
	\section{Introduction}
	
The role of mass loss from massive stars, especially episodic mass loss
in evolved massive stars, is one of the outstanding open questions
facing stellar evolution theory \citep{Smith2014}. While the upper limit
to the masses of stars is thought to be 150~\msun
\citep{Figer2005,Oey2005}, and was even claimed to exceed 300~\msun
\citep{Crowther2010, Banerjee2012}, the masses of hydrogen-deficient
Wolf-Rayet (WR) stars do not exceed 20~\msun
\citep{Crowther2007}. Classical line-driven wind theory
\citep{Kudritzki2000}, once thought to be responsible for removing the
envelopes of massive stars, has been shown inadequate, both on
theoretical grounds \citep[due to clumping,][]{Owocki1999} and
estimations based on spectral lines
\citep{Bouret2005,Fullerton2006,Cohen2014}, which demand reductions in
the mass-loss rates by a factor of $\sim$2-3. So how do massive stars
shed their envelopes? Binary interactions via Roche-Lobe overflow (RLOF)
are predicted to occur in 70\% of massive stars and strip the envelopes
in $\sim30\%$ of O stars, given the high binarity fraction ($\sim70\%$)
of massive stars \citep{Sana2012}. Episodic mass loss is possibly the
dominant process that operates in single stars, however, the physical
mechanism responsible remains a mystery \citep{Smith2014}.

The importance of episodic mass loss has come to the forefront in both the massive star and supernova (SN) communities. \textit{Spitzer} images have revealed numerous circumstellar shells surrounding massive, evolved stars in our Galaxy \citep{Gvaramadze2010,Wachter2010}. Episodes of enhanced mass loss have been recorded not only in luminous blue variables (LBVs), but also in extreme red supergiants \citep[RSGs, e.g. VY CMa,][]{Decin2006}. Moreover, untargeted supernova surveys have found dusty circumstellar material around superluminous supernovae \citep[SLSN,][]{Gal-Yam2012}, and mysterious optical transients with luminosities intermediate between novae and supernovae. The presence of circumstellar material implies a central role of episodic mass loss in the evolution of massive stars and this proposal aims to confirm this hypothesis. Tantalizing evidence suggests that SLSN occur in low-metallicity host galaxies \citep{Neill2011}, implying that such supernovae dominated the metal-poor early Universe. The overluminous Type IIn SN 2010jl is a well-studied example of a SLSN, with a massive progenitor star (30 \msun) surrounded by a dense circumstellar shell \citep{Smith2011b,Zhang2012}, which exploded in a low-metallicity galaxy \citep{Stoll2011}. SN2008S, a well-studied example of the class of intermediate-luminosity optical transients, was found to have a dust-enshrouded progenitor \citep[8--10 \msun,][]{Prieto2008} in pre-explosion \textit{Spitzer} images of the host galaxy NGC 300. Finally, the remarkable SN2009ip involves a 50--80 \msun\, progenitor that underwent a series of episodic mass loss events. Its spectacular finale included a series of eruptions in 2009 and 2010 until its final explosion in 2012 as a Type IIn supernova \citep{Mauerhan2013}, although this was contested \citep{Pastorello2013,Fraser2013}. These examples strongly suggest that episodic mass loss in massive stars is central to their evolution and therefore has profound consequences for the enrichment of the interstellar medium and the chemical evolution of the early Universe.

The physics of LBV eruptions, pre-SN eruptions and extreme RSG mass-loss is still in its infancy and, as stated in the review by \citet{Smith2014}, “is a major unsolved problem in astrophysics”. Models of single-star evolution adopt empirical, constant mass-loss prescriptions, which highly influence the outcome \citep{Meynet2015}. The ASSESS project tackles the role of episodic mass loss in massive stars by using the fact that mass-losing stars form dust and are bright in the mid-infrared (mid-IR). Physically, there are a number of ways a massive star can become a source of significant mid-IR emission. First, dust can form in a dense, but relatively steady stellar wind. In the most extreme cases, such as in the progenitors of the SN 2008S and the NGC300-OT 2008 transient \citep{Bond2009}, the wind is optically thick even in the near-IR and the source star is seen only in the mid-IR \citep{Prieto2008}. Second, a very massive star can have an impulsive mass ejection or eruption with dust forming in the ejected shell of material. Initially the optical depth and dust temperatures are high, but then drop as the shell expands. The most famous example is the “great eruption” of $\eta$ Carinae in the 19th century \citep{Humphreys1994, Davidson1997, Smith2011a}, which ejected several solar masses of material. Third, the dust can be located in a circumstellar disk and emit over a broad range of temperatures, as is seen in supergiant B[e] stars (sgB[e]) stars \citep{Zickgraf2006}.

While stars with significant mid-IR emission are intrinsically rare, many of the most interesting massive “superstars”, such as $\eta$ Car or “Object X” in M33 \citep{Khan2011}, belong to this class. It is clear that searching for analogs of these interesting stars using mid-IR photometry of nearby galaxies is the way to go. The existing mid-IR “roadmaps” for interpreting luminous massive stars \citep{Bonanos2009,Bonanos2010} are based on known massive stars in the LMC and the SMC. They have identified LBVs, sgB[e], and RSGs among the brightest mid-IR sources, due to their intrinsic brightness and due to being surrounded by their own dust. What is new about ASSESS is the idea of conducting – for the first time – a systematic study of mass loss in massive stars, by selecting targets using mid-IR photometry of nearby galaxies obtained with \textit{Spitzer}.

	\section{Methodology}

We have collected recently published mid-IR photometric catalogs from \textit{Spitzer} of galaxies with high star-formation rates within 5 Mpc: (a) seven dwarf galaxies within 1.5 Mpc from the DUSTiNGS project \citep{Boyer2015}: IC 10, IC 1613, Phoenix, Pegasus, Sextans A, Sextans B, and WLM, (b) 13 galaxies within 5 Mpc \citep{Khan2015,Khan2017}: M31, M33, NGC 247, NGC 300, NGC 1313, NGC 2403, M81, M83, NGC 3077, NGC 4736, NGC 4826, NGC 6822, and NGC 7793, and (c) five galaxies within 4 Mpc \citep{Williams2016}: NGC 55, NGC 253, NGC 2366, NGC 4214, and NGC 5253. The mid-IR photometry made available by the SAGE surveys of the LMC \citep{Meixner2006} and SMC \citep{Gordon2011} has been also searched for undetected, dust-obscured targets in our nearest neighbor galaxies. These catalogs contain mid-IR photometry of over 5 million point sources in 27 nearby galaxies, 19 of which have Pan-STARRS1 coverage \citep{Chambers2016}, providing an ideal dataset for a systematic study of luminous, dusty, evolved massive stars. We have compiled mid-IR photometric catalogs for these galaxies, including their counterparts in Pan-STARRS1 ($g,r,i,z,y-$bands), 2MASS \citep{Cutri2003}, VISTA Science Archive, WISE \citep{Cutri2012} and other archival surveys of particular galaxies to construct their spectral energy distributions (SEDs) out to 24 $\mu$m. The single epoch $5\sigma$ depth of Pan-STARRS1 ranges from 22nd magnitude in $g-$band to 20th magnitude in $y-$band, corresponding to absolute magnitudes brighter than $-$6 in $g$ and $-$8 in $y$ at 3.5 Mpc, respectively, which include the most luminous, evolved targets. 

Based on these catalogs, we have selected over 1000 luminous and red sources (selected by their colors in $[3.6]-[4.5$]) in these 27 galaxies and are conducting follow-up low-resolution spectroscopy of these sources, mainly with FORS2 on VLT and OSIRIS on GTC. The spectra yield stellar types, luminosity classes, effective temperatures and an estimate of the reddening. High-resolution spectra will be obtained for particularly interesting targets for further analysis. 

SED modeling with DUSTY \citep{Ivezic1997} will provide radii and age estimates of the circumstellar shell, as well as the dust temperature, ejected mass, and bolometric luminosity. SED shapes will be quantified to estimate the timescales of episodic mass loss and lifetimes of the various evolved stages as a function of spectral type and metallicity. Evidence of binarity (from spectra, SEDs, light curves) will provide an estimation of the relative contribution of RLOF to the observed dusty evolved stages of massive stars. Armed with all these parameters for a sample of $\sim~1000$ dusty, evolved stars, spanning a range of metallicity ($\sim$1/15 $-$ 2\,\zsun), we will perform a comparison with state-of-the-art stellar evolutionary models \citep{Brott2011,Ekstrom2012,Georgy2013b, Meynet2015} to evaluate the input mass-loss rates and predicted outcomes. We plan to reverse-engineer the target stars to quantify and confirm the amount of “input” episodic mass loss needed to match the measurements.

	\section{Results}

\subsection{Photometric Classifier}

We have employed state-of-the-art machine-learning algorithms to automatically classify and select types of mass-losing stars, thereby accelerating and systematizing the investigation of multi-wavelength photometry. We developed a classifier for evolved massive stars based on known massive stars in M31 and M33 and using color indices as features to classify evolved massive stars into the following categories:
blue, yellow, red supergiants, LBVs, classical Wolf-Rayet stars, sgB[e]. We also included a class for outliers (e.g. background galaxies, AGNs). The classifier is found to be on average 83\% accurate \citep{Maravelias2022}. We are currently applying this classifier to classify over one million sources in 25 nearby galaxies (see Maravelias et al., this volume).

\subsection{Observational survey}

We have prioritized our targets based on their luminosity and IR excess, specifically, giving highest priority to targets with $m_{3.6}-m_{4.5} \geq 0.5$ mag and $M_{3.6} \leq - 9.75$ mag. We have obtained multi-object spectroscopy with both the VLT and GTC starting in 2020, giving priority to the galaxies that had enough high-priority targets to justify multi-object spectroscopy. We used the FORS2 spectrograph (Tramper et al., in prep.) and obtained spectra of over 400 high-priority and over 500 "filler" stars in M83, NGC~55, NGC~247, NGC~253, NGC~300, NGC~7793, Sextans~A and WLM over 43h. The spectra have a resolving power of $R=1000$ and a wavelength coverage around $5400-8200$ \AA, which is suitable for classification and parameter estimation. Figure 1 shows examples of dusty, evolved massive stars identified in NGC~55, NGC~300 and NGC~7793.\\

	\begin{figure}[ht]
		\begin{center}
			\includegraphics[width=4.1in]{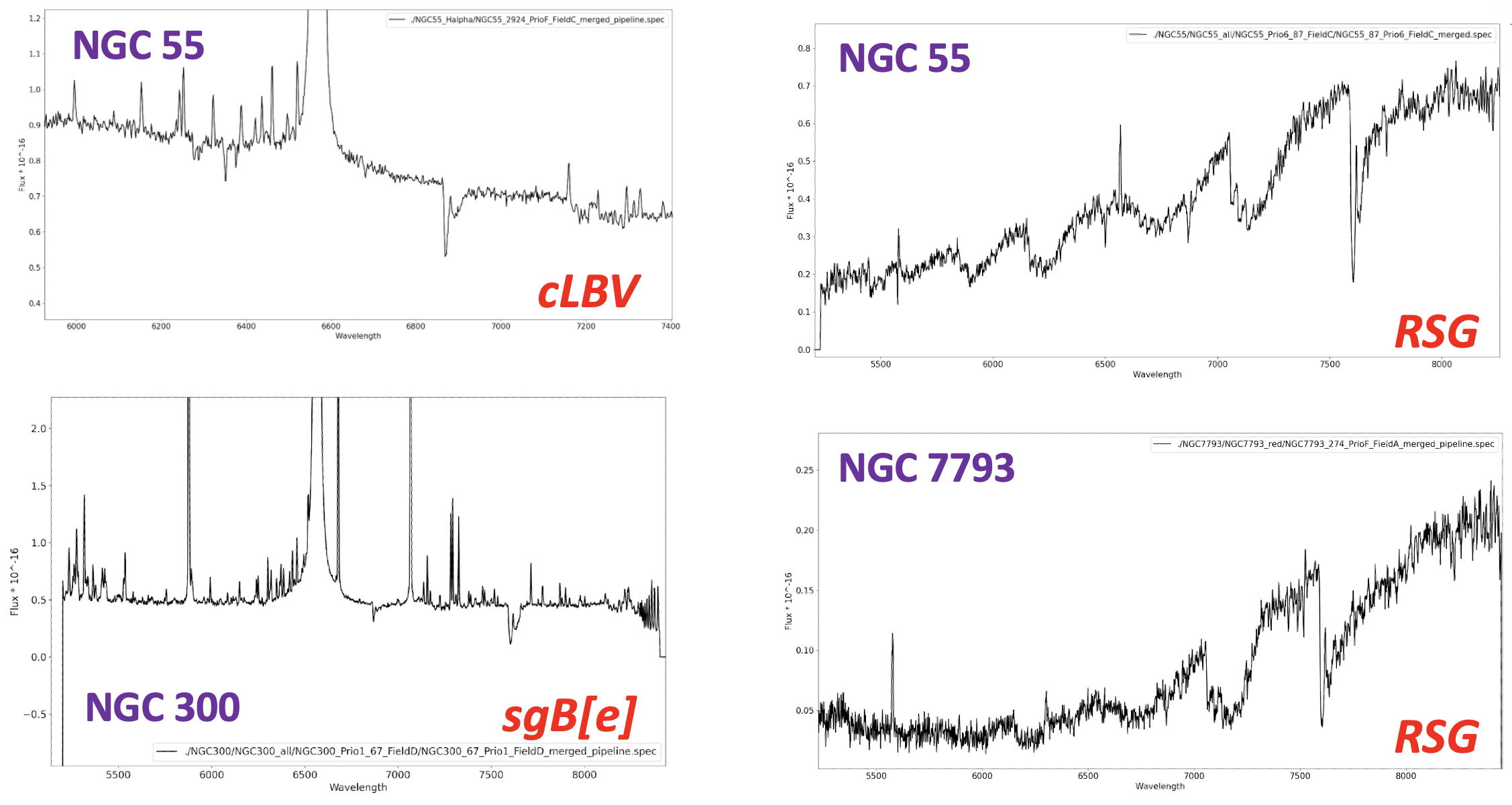} 
		\end{center}
			\caption{Newly identified spectra of evolved massive stars in our FORS2 data from the VLT, including a candidate LBV and RSG in NGC~55 (top row), a sgB[e] in NGC~300 and a RSG in NGC~7793 (bottom row).}
	\end{figure}

We also used the GTC OSIRIS spectrograph (for details see Munoz-Sanchez et al., this volume) and have so far obtained 48 high-priority stars in NGC~6822 and 33 in IC~10. The GTC spectra have a resolving power of $R\sim500-700$ and a wavelength coverage around $5200-9200$ \AA, which are being used to classify the sources and obtain their parameters.

In the Magellanic Clouds, we have similarly selected dusty, evolved
sources and obtained spectra with the MagE spectrograph on Magellan and
identified 8 new RSGs. Among them is a luminous, extreme RSG, with
similar properties to WOH G64. Our results are presented by de Wit et
al. (2022, this volume).

\subsection{Mass loss rates}

We have set out to determine the mass loss rates (MLR) of red supergiants in the Small Magellanic Cloud, based on the catalogs of \citet{Yang2020} and \citet{Ren2021}. Comprehensive photometry in over 50 bands (from the UV to 24$\mu m$) for over 2000 RSG has been compiled and a grid of DUSTY models \citep{Ivezic1997} was created for both silicate and carbon dust. This grid was used to perform a $\chi ^2$ fit of the SEDs and determine the dust parameters, optical depth and the mass loss rate for each supergiant. 

From the distribution of MLR, we find a typical value of $\sim 10^{-6}\, \msun\, yr^{-1}$, with a few outliers at around $\sim 10^{-4}$ and $10^{-3}\, \msun\, yr^{-1}$. We determine a new MLR vs. $L$ relation based on an unbiased sample of RSG in the SMC, which shows an upturn at around $\log (L/\lsun)= 4.7$, with enhanced mass loss occurring at higher $L$. Compared to previously determined relations in the SMC, our result (Yang et al. 2022, in prep.) is most similar to the relations of \citet{Feast1992} and \citet{vanLoon2005}. We plan to apply the procedure to all our program galaxies and determine the MLR at a range of metallicities. 

\section{Conclusions}

We have presented the first results of an ambitious systematic study of episodic mass loss in $\sim 1000$ evolved massive stars. This survey is timely, given the recent availability of mid-IR catalogs and ambitious, as it plans to increase the number of evolved massive stars in nearby galaxies by a factor of 5. The \textit{James Webb Space Telescope} is operating concurrently with this project. The enormous boost in sensitivity and angular resolution will revolutionize our understanding of these nearby objects. However, to fully exploit this we need to be able to tie the JWST results into the more general population. This project provides this anchor.

The results of this study will not only provide the first quantitative inventory and characterization of dusty massive stars in 27 galaxies in the nearby Universe at a range of metallicities, but may also reveal new classes of enshrouded stars and rare transitional objects. A byproduct of the survey will be the release of multi-wavelength photometric catalogs of luminous sources in 27 galaxies, including their classifications, which will be valuable for various scientific projects.

\section*{Acknowledgments}

We acknowledge funding support from the European Research Council (ERC) under the European Union’s Horizon 2020 research and innovation programme (Grant agreement No. 772086).

\bibliographystyle{apj}
\bibliography{./references}

\begin{discussion}

\discuss{Paul Crowther}{Could some of the low-luminosity sources be AGB stars?}

\discuss{Alceste Bonanos}{We do not think we have much AGB contamination, as we have used conservative selection criteria based on the work of Yang et al. (2020).}

\discuss{Alex de Koter}{We expect RSGs to predominantly have Oxygen (i.e. Si) dust. What are the luminosities of the sources that needed carbon dust to be fitted?}

\discuss{Alceste Bonanos}{I have to check this. However, we first modeled all sources with Si dust and only proceeded with a carbon dust fit for those that did not fit well with Si dust. We don't expect to have much AGB contamination.}

\discuss{Floor Broekgarden}{Why were the stars in your survey quite different compared to Emily Levesque? Especially in the HR diagram Levesque seemed to find more low T systems. Is this just because of the difference in surveys?}

\discuss{Alceste Bonanos}{We used mid-IR selection criteria, so we expect to find more extreme RSGs.}

\discuss{Pat Morris}{Following the question by Alex de Koter about finding silicate dust in the SMC RSG fits to be a little suspicious, more likely in lower mass (luminosity) AGB stars.  Might this be an effect of more than one thermal component? I did not catch whether this was considered in the fits, might also help with the poorer fits you showed.}

\discuss{Alceste Bonanos}{We have not modeled more than one thermal component, we should try this.}

\end{discussion}
	
\end{document}